\documentclass[dvips]{article}
\usepackage{icrctc07}

\title{TeV and X-ray Monitoring of LS I +61 303 With VERITAS, Swift, and RXTE}
\shorttitle{TeV and X-ray Monitoring of LS I +61 303 With VERITAS, Swift, and RXTE}

\authors{A. Smith$^{1,2}$for the VERITAS collaboration$^{3}$ }
\shortauthors{Author and et al.}
\afiliations{$^1$Smithsonian Astrophysical Observatory\\ $^2$University of Leeds\\$^3$ For a complete author list please see G. Maier et al., ``VERITAS: Status and Latest Results'', these conference preceedings.}
\email{smith@egret.sao.arizona.edu}

\abstract{ Between September 2006 and February 2007, the galactic binary LS I +61 303 was monitored in the TeV band with the VERITAS array of imaging Cherenkov telescopes. These observations confirm LS I +61 303 as a variable TeV gamma-ray source, with emission peaking between orbital phase 0.6 and 0.7. During this observational period, monitoring in the X-ray regime was also carried out using both the RXTE and Swift detectors, which offered complementary coverage of the source. Outbursts in the 0.2-10 keV band were observed by both satellites at close to the same orbital phase as the TeV peak during the 2 orbital cycles covered simultaneously in both bands. While this source has been extensively studied in the X-ray band in the past, this is the first observational campaign to utilize contemporaneous X-ray and TeV data on LS I +61 303.}

\begin{document}
\maketitle
\section{Introduction}
\begin{figure*}
\begin{center}
\includegraphics[width=\textwidth, height=80mm]{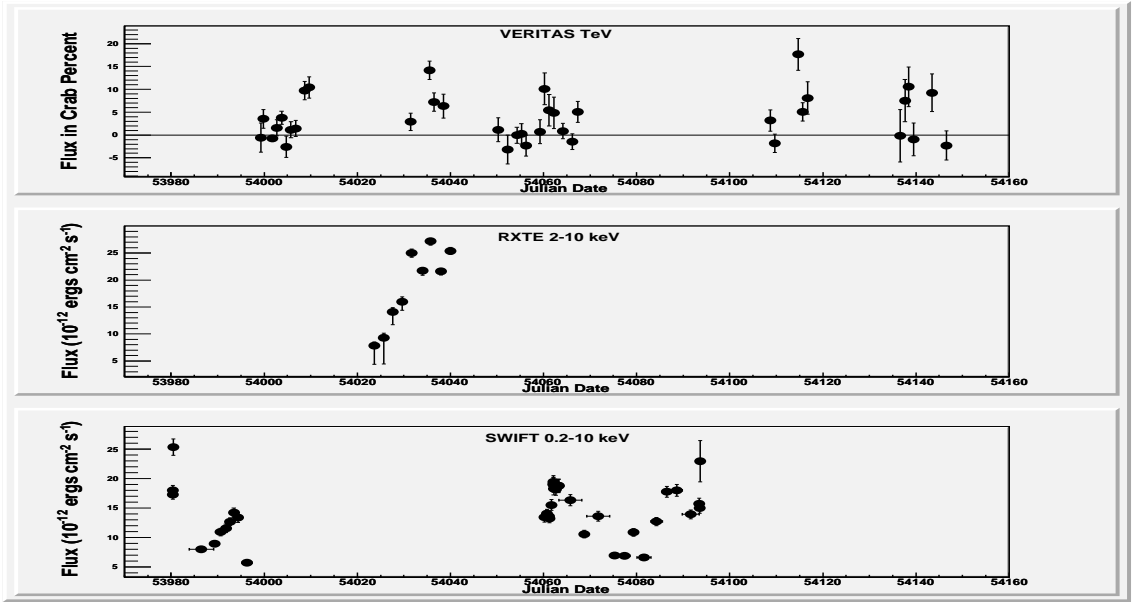}
\end{center}
\caption{VERITAS, RXTE, and Swift monitoring of LS I +61 303 over seven orbital periods. Peaks can be seen in all three data sets near dates corresponding to $\phi$=0.7 (also see figure 2).}\label{fig1}
\end{figure*}
First associated with the COS-B source 2CG 135+01 \cite{COSB} in 1978, LS I +61 303 has been a source of considerable interest for the last 3 decades due to its peculiar behavior in radio and X-rays. Located at $\sim$2 kpc distance \cite{Casares}, LS I +61 303 is a pairing of a Be star with a neutron star or a black hole which completes an orbital transit every 26.496 days \cite{LSIperiod}. Flaring behavior has been historically detected in radio\cite{LSIradio}, X-rays\cite{firstRXTE}\cite{secondRXTE}, and most recently TeV gamma rays\cite{Albert}. X-ray campaigns show flaring activity occuring regularly, every orbital cycle around orbital phase 0.4$\rightarrow$0.7 \cite{firstRXTE}\cite{secondRXTE}\cite{ROSAT1}\cite{ROSAT2}. The most interesting feature of the X-ray flares is that they consistently appear to precede the radio outbursts which occur several days later. LS I +61 303 has also been associated with the EGRET GeV source 3EG J0241\cite{EGRETLSI} which showed evidence for outbursts both near perisastron passage ($\sim$0.23\cite{LSIperiod}) and later in the orbit near phase 0.5 \cite{MassiEGRET}.

Traditionally, there have been two classes of models surrounding variable high-energy emission from this source. The first \cite{microquasarmodel}\cite{Massijet} assumes that the broadband emission is generated by accretion around the neutron star as it encounters varying stellar wind densities in its orbit (microquasar models). Higher accretion rates cause a radio jet to form which can then upscatter stellar photons to TeV energies. Observations resolving compact one-sided radio emission from the system were taken as unambiguous evidence for the presence of a relativistic jet and thus proof of the microquasar model \cite{Massijet}. 

Alternatively, in the models of \cite{initialpulsar}\cite{Dubus2006}\cite{Onion}, the system is treated as a binary pulsar with particle acceleration taking place in the shock formed between the pulsar and stellar winds. These pulsar models interpret the resolved radio emission in \cite{Massijet} as coming from the cometary tail of the pulsar wind as it is blown about by the stellar wind. This assertion is lent further credence by the more in-depth radio imaging of this cometary tail detailed in \cite{Dhawan}. Although the binary pulsar model for LS I +61 303 seems much more likely in light of these recent observations, there are still unresolved issues (see \cite{Romero}) in both models which need further multiwavelength observations to investigate.

\section{Observations}
\subsection{VERITAS TeV Observations}
Sensitive in the energy range 100 GeV to 30 TeV, VERITAS is an newly commissioned array of four 12-m imaging Cherenkov telescopes which has begun its full observational program as of January 2007 (see G. Maier in these proceedings). During the Fall of 2006, the initial 2-telescope stereo array was in operation, with a third telescope being added for observations from January onward. TeV rates quoted henceforth are in terms of Crab Nebula fluxes taken from contemporaneous observations (see O.Celik in these proceedings).

From September 2006 until February 2007, LS I +61 303 was observed for a total of 43.3 hours covering orbital phases 0.2$\rightarrow$0.9. These observations covered 5 orbital cycles with reliable detections resulting in every orbital period with the exception of the February dark run (see G. Maier in these conference proceedings). However, the data taken in February were from a shorter exposure and are consistent with the behavior seen in the other months. During the ``OFF" phases, i.e., 0.2$\rightarrow$0.5 and 0.8$\rightarrow$0.9, the source is quiescent in the TeV range, with no reliable detection resulting. During the ``ON'' phase, 0.5$\rightarrow$0.8, VHE gamma rays are detected with emission ranging between 4$\%$ and 13$\%$ of the Crab Nebula flux. VERITAS observations showed the TeV flux peak centered at around 0.65$\rightarrow$0.74, which is later in phase than the TeV maximum observed by MAGIC \cite{Albert} at 0.5$\rightarrow$0.6. 
\begin{figure*}
\begin{center}
\noindent
\includegraphics [width=\textwidth, height=80mm]{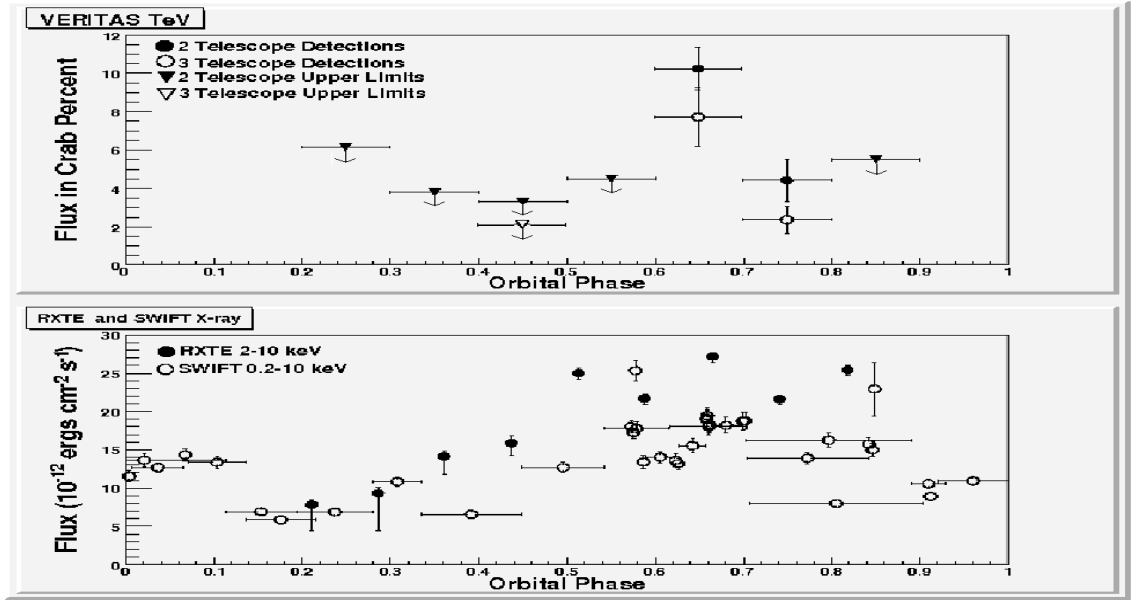}
\end{center}
\caption{Comparisons between contemporaneous TeV and X-ray data taken with VERITAS, RXTE, and Swift. TeV detections/upper limits are $\>$3$\sigma$/99.9$\%$ Helene upper limits. All three observation sets show elevated emission between phases 0.5 and 0.8.}\label{fig2}
\end{figure*}

\subsection{RXTE Observations}
Monitoring in the 2-10 keV band was carried out by RXTE \cite{RXTE} for nine 1-ks observations from 2006/10/15 to 2006/10/31. Spectral fitting was carried out with XSPEC 5.12.3.1 assuming an absorbed power law model. Fluxes shown in figures 1 and 2 come from integrating this model over the 2-10 keV range and are scaled to emission seen by RXTE from the Crab Nebula \cite{RXTEcrabflux}. The light curve produced from this fitting shows a clear flare evolving from around orbital phase 0.45, peaking at around 0.7 and persisting until the end of the data set at 0.8. This flare was particularly strong, with emission peaking at 27$\times$10$^{-12}$ergs s$^{-1}$cm$^{-2}$ which is larger than, but consistent with, past RXTE observations of this source\cite{firstRXTE}. The spectra fit to each data point show no strong evidence for spectral variation, however, the relatively short exposure time and reduced instrument operation (i.e. only one PCU on for some nights) must be taken into account. A more detailed description of these observations will be provided in an upcoming publication.

\subsection{Swift Observations}
From September 2006 to October 2006, and again in November 2006-January 2007, the XRT instrument aboard the Swift satellite \cite{SWIFT} was used to monitor LS I +61 303 in the 0.2-10 keV band. Reduction and analysis procedure can be found in Holder, Falcone and Morris (this conference). The XRT count rate to flux conversion factor used for this data was
5.7$\times$10$^{-11}$ (erg/cm${^2}$/s)/(c/s).  This conversion factor was obtained using
the mean spectral fit parameters for these data.  The systematic error associated with spectral variation is not included in the error bars. This data set overlapped only once with VERITAS TeV data but fortunately it covered a region where VERITAS saw a clear detection of the source. Similar to the RXTE observations, the Swift data set shows a clear elevation of emission occuring between orbital phases 0.5 and 0.8. The data set also shows a secondary elevated emission occurring between orbital phases 0.0 and 0.1. 

\section{Results}
 \subsection{TeV Variability}
It can be seen from figures 1 and 2 that the phase of the peak emission in the VERITAS data is always near the phase region 0.65$\rightarrow$0.7. The VERITAS data have a maximum flux between 0.65 and 0.75, which, if compared to the location of the reported MAGIC TeV maximum of 0.5 to 0.6,  appears to be slightly offset in phase. However, a full statistical analysis of the light curves has yet to be carried out. Also, both the MAGIC and VERITAS light curves do not evenly sample the entire orbital phase.  More long-term, evenly sampled observations are needed to investigate the nature of the TeV variability from the LS I +61 303.

\subsection{Correlation Between Data Sets}

Shown in figure 2 is a comparison between the TeV and X-ray data binned by orbital phase. It appears that there may be a correlation between the rise time of TeV and X-ray emission in the system, although a full statistical analysis has yet to be carried out. Further analysis of the data as well as further long-term simultaneous observations are necessary to investigate this correlation further.

\section{Discussion}
With the preliminary analysis presented here we are unable to make any
firm statements constraining either the microquasar or the binary pulsar
models. It is noted that the X-ray and TeV emission appear to be correlated in time. Further analysis of this data set will be performed to study the nature of a possible correlation between the collected data from RXTE, Swift, and VERITAS. This analysis will be presented in an upcoming publication.

\subsection*{Acknowledgments}

This research is supported by grants from the U.S. Department of Energy, the
U.S. National Science Foundation and the Smithsonian Institution, by NSERC in
Canada, by PPARC in the U.K. and by Science Foundation Ireland.

\bibliographystyle{unsrt}
\bibliography{icrc.bib}
\end{document}